\documentclass[graybox]{svmult}

\usepackage{type1cm}            % activate if the above 3 fonts are
\usepackage{makeidx}            % allows index generation
\usepackage{graphicx}           % standard LaTeX graphics tool
\usepackage{multicol}           % used for the two-column index
\usepackage[bottom]{footmisc}   % places footnotes at page bottom

\usepackage{newtxtext}
\usepackage[varvw]{newtxmath}
\usepackage{acronym}
\usepackage[numbers,sort&compress]{natbib}
\usepackage{amsmath,amsfonts}
\usepackage{psfrag,epsf}
\usepackage{enumerate}
\usepackage{url}
\usepackage{tabularx}
\usepackage{geometry}
\usepackage{dirtytalk}
\usepackage[T1]{fontenc}
\usepackage[utf8]{inputenc}
\usepackage{authblk}

\usepackage{booktabs}
\usepackage[scale=2]{ccicons}

\usepackage{xspace}

\usepackage[colorlinks=true,
            linkcolor = gray,
            citecolor = gray]{hyperref}

\DeclareRobustCommand{\rpsi}{\text{\reflectbox{$\psi$}}}

\makeindex             % used for the subject index
\begin{document}

\title*{In search of the perfect fit: interpretation, flexible modelling, and the existing generalisations of the normal distribution}
\titlerunning{The perfect fit: interpretable flexible modelling}
\author{Andriette Bekker, Matthias Wagener, and Mohammad Arashi}
\institute{Andriette Bekker
\at Department of Statistics, University of Pretoria, 0002, South Africa; South Africa and Centre of Excellence in Mathematical and Statistical Sciences, Johannesburg 2000, South Africa
\email{andriette.bekker@up.ac.za}
\and
Matthias Wagener
\at Department of Statistics, University of Pretoria, Pretoria, 0002
\email{matthias@dilectum.co.za}
\and
Mohammad Arashi
\at Department of Statistics, Faculty of Mathematical Sciences, Ferdowsi university of Mashhad, Iran
\email {arashi@um.ac.ir}
\at Department of Statistics, University of Pretoria, Pretoria, 0002}
\maketitle
\abstract{Many generalised distributions exist for modelling data with vastly diverse characteristics. However, very few of these generalisations of the normal distribution have shape parameters with clear roles that determine, for instance, skewness and tail shape. In this chapter, we review existing skewing mechanisms and their properties in detail. Using the knowledge acquired, we add a skewness parameter to the body-tail generalised normal distribution \cite{BTGN}, that yields the \ac{FIN} with parameters for location, scale, body-shape, skewness, and tail weight. Basic statistical properties of the \ac{FIN} are provided, such as the \ac{PDF}, cumulative distribution function, moments, and likelihood equations. Additionally, the \ac{FIN} \ac{PDF} is extended to a multivariate setting using a student t-copula, yielding the \ac{MFIN}. The \ac{MFIN} is applied to stock returns data, where it outperforms the t-copula multivariate generalised hyperbolic, Azzalini skew-t, hyperbolic, and normal inverse Gaussian distributions.
}
\vspace{2cm}
\subsection*{List of abbreviations}
\small
\begin{acronym}[NASDAQ] % Give the longest label here so that the list is nicely aligned
\acro{FIN}{flexible and interpretable normal distribution}
\acro{MFIN}{multivariate flexible and interpretable normal distribution}
\acro{PDF}{probability density function}
\acro{CDF}{cumulative distribution function}
\acro{EP}{exponential power}
\acro{BTN}{body-tail generalised normal}
%\acro{MGF}{moment generating function}
\acro{FTN}{flexible tail normal}
\acro{LL}{log-likelihood}
\acro{ML}{maximum likelihood}
\acro{HYP}{hyperbolic}
\acro{GHYP}{generalised hyperbolic}
\acro{NIG}{normal inverse gaussian}
\acro{ST}{Azzalini skew-t distributions}
\acro{AIC}{Akaike information criterion}
\acro{ACST}{Acasti Pharma}
\acro{GSIT}{GSI Technology}
\acro{LOAN}{Manhattan Bridge Capital}
\acro{MGI}{Moneygram International}
\acro{PSCM}{PowerShares S\&P SmallCap Materials Portfolio}
\acro{NASDAQ}{National Association of Securities Dealers Automated Quotations}
\acro{LHS}{left-hand side}
\acro{RHS}{right-hand side}
\end{acronym}
\normalsize
\pagebreak

\section{Introduction}
\label{sec:intro}
Flexible modelling is a branch of distribution theory that has been evolving since 1879, when Galton proposed the log-normal distribution \cite{deVries1894}. Today, further advancements have led to a greater appreciation for the importance of flexible modelling, with more researchers focusing on this area. As a result, there has been a surge of creativity and innovation in the field as experts discover new distributions and develop more effective methods for using them. Some of the more prominent flexible distributions include:
\begin{itemize}
    \item finite mixture models \cite{mclachlan2000}.
    \item variance-mean mixtures \cite{barndorff82}.
    \item copulas \cite{nelsen07}.
    \item Box-Cox transformation \cite{box64}.
    \item order-statistics-based distributions \cite{jones04}.
    \item probability integral transformations of \cite{ferreira06}.
    \item Pearson system of distributions \cite{johnson94}.
\end{itemize}
New models today don't just require a better fit, they demand a wider focus than simply fitting better. For new models to be successful, it is important to consider a wide range of desirable qualities. As discussed by the authors in \cite{ley14flex, jones15, mcleish1982robust, punzo2021multivariate, BTGN, ley2021flexible}, these qualities include:
\begin{itemize}
	\item A manageable number of easily interpretable parameters, such as those controlling location, scale, skewness, and kurtosis).
	\item Suitable estimation properties that enable good predictions and inferences that could be used for tests of normality.
	\item Mathematical tractability with simple formulae that aid in exposition, improve computational power, and augment speed.
\end{itemize}
This chapter focuses on a particular generalisation of the normal distribution, that has been referred to by a variety of names, such as the \ac{EP}, generalised power, generalised error, generalised Gaussian, and generalised normal distribution. This family of distributions was first suggested by \cite{subbotin23}, and then again by both \cite{box62} and \cite{box73}. The \ac{EP} \ac{PDF} is given below:
\begin{equation}
 \label{eq:gen_den}
 \phi(z) = \frac{s}{2\Gamma(\frac{1}{s})}\textrm{e}^{-\left|z\right|^{s} \notag},
\end{equation}
where $z\in\mathbb{R},s>0$. The \ac{EP} distribution, that is symmetric by nature, has been extended in a variety of ways to account for skewness in the data. The goal of this chapter is to introduce the flexible and interpretable normal distribution (\ac{FIN}). By studying the existing research on skewing mechanisms, 
we can gain an understanding of how skewing mechanisms function and what properties they have. We start with Table \ref{tab:PNskewtimeline}, which gives a summary of the existing skew EP generalisations and their characteristics. From the latter, we see that two-piece scaling appears to have good overall characteristics, but it depends on the data whether the skewness originates from an asymmetry in scale or an asymmetry in kurtosis, prompting the development of skewing with kurtosis. The Azzalini \ac{PDF} perturbation introduces skewness by reducing the tail weight of one tail without increasing the weight of the other. The Azzalini PDF perturbation affects the central tendency of the distribution in a manner where neither the mean, mode, nor median is equal to the estimated location parameter. Furthermore, the mode of the distribution must be computed numerically \cite{jones2003skew}. The two-piece kurtosis skewing addresses this limitation by allowing the possibility of decreasing tail weight for one half of the distribution, although the other half cannot have more tail weight than the original distribution. This approach may lead to inelegant body shapes, such as having a normal left half and exponential right half with an obvious ``joint" at the mode. Although this approach may be necessary in certain cases of data modelling, it can introduce issues with the \ac{LL} function due to non-differentiable density at the mode, see \cite{zhu2009properties}. Table \ref{tab:skewmech} provides a summary of general skewing mechanisms and their traits. The Jones skew t is a much neglected distribution that is elegant in its approach to adding skewness through asymmetric kurtosis. By varying degrees of freedom asymmetrically, one half can have less tail weight than the other without limitation. The Jones skew t also does not have an obvious ``joint" at the mode because the body shape of both sides stays the same. The drawback of this approach is that the t distribution cannot accommodate lighter tails than normal, and moments are not defined for very low degrees of freedom and high degrees of skewness. The sinh-arcsinh transformation is another method for skewing with desirable properties, but it is not suitable for the \ac{EP} distribution since the transformation distorts the body shape of the baseline distribution, see \cite[pg. 763, fig. 1]{jones09sinh}. This leads to the body shape parameter having a confounded interpretation relative to the baseline \ac{EP} distribution. The remaining skewing mechanisms in Table \ref{tab:skewmech} can easily add many parameters, but they may lack interpretation.
\begin{table}[ht!]
\caption{Timeline of derived skewed \ac{EP} distributions and their characteristics.}
\label{tab:PNskewtimeline}
\fontsize{8}{4}\selectfont%\footnotesize
\centering
 \begin{tabularx}{0.9\linewidth}{XXXXXXXX}
        \addlinespace[4pt]
        \hline
        \hline\addlinespace[4pt]
        Distribution & Year & Extension method & Parameter count\footnotemark & Interpretable parameters\footnotemark & Closed form \ac{PDF} & Trouble free estimation & Parsimonious \\\addlinespace[4pt] \hline\addlinespace[4pt]
         Exponential power \cite{box73,ayebo03,Komunjer2007} & 1973, 1995, 2003, 2007 & Two-piece scale& 4 & Yes & Yes & Yes & Yes \\\addlinespace[4pt] \hline\addlinespace[4pt]
        Azzalini skew GED \cite{azzalini1986further,bekker2018computational} & 1986 & Azzalini \ac{PDF} perturbation  & 4 & Partially & Yes & No & Yes 
         \\\addlinespace[4pt] \hline\addlinespace[4pt]
        Asymmetric exponential power \cite{zhu2009properties,bottazzi11} & 2009, 2011 & Two-piece shape skewing & 5 & Yes & Yes & No & No \\\addlinespace[4pt] \hline\addlinespace[4pt]
        BGN SAR \cite{cintra2014beta} & 2014 & Beta generator & 5 & No & Yes & Yes & No  \\\addlinespace[4pt] \hline\addlinespace[4pt]
        Tukey’s g-h GED \cite{jimenez2015generalization} & 2014 & g-h transformation \cite{tukey1977exploratory} & 5 & No & No & No & No \\\addlinespace[4pt] \hline\addlinespace[4pt]
        Flexible-skew \ac{EP} \cite{yilmaz2016flexible} & 2016 & Extended Azzalini type \ac{PDF} perturbation  & 6 & No & Yes & Yes & No
        \\\addlinespace[4pt] \hline\addlinespace[4pt]
        Alpha-skew generalised normal \cite{Altun2019,mahmoudi2019alpha} & 2019 & Elal-Olivero \ac{PDF} perturbation \cite{elal2010alpha} & 4 & No & Yes & Yes & No \\\addlinespace[4pt] \hline\addlinespace[4pt]
\end{tabularx}
\end{table}
\footnotetext[1]{Including location and scale.}
\footnotetext[2]{Existing parameterisations that include interpretations for location (equivalent to mean or mode), scale, skewness, and kurtosis.}
\footnotetext[3]{In the case of a normal baseline distribution.}
\begin{table}[h!]
\caption{Timeline of existing skewing mechanisms for distributions and their characteristics.}
\scriptsize
\centering
\fontsize{8}{2}\selectfont%\footnotesize
\begin{tabularx}{0.9\linewidth}{XcXcccc}
%    \begin{tabular}{l c c c c c c}
        \addlinespace[4pt]
        \hline
        \hline\addlinespace[4pt]
Family or distribution & Year & Extension method& Parameter count$^1$ & Interpretable parameters$^2$ & Flexible tail$^3$ & Parsimonious \\\addlinespace[4pt] \hline\addlinespace[4pt]
        Two-piece normal
        \cite{fechner1897,mudholkar2000epsilon}
         & 1897 & Two-piece scaling & 3 & Yes & No & Yes \\\addlinespace[4pt] \hline\addlinespace[4pt]
         Azzalini \cite{azzalini1985} & 1985 & \ac{PDF} perturbation & 3 & Partially & No & Yes
         \\\addlinespace[4pt] \hline\addlinespace[4pt]
         Balakrishnan skew normal
         \cite{arnold2002skewed} & 2002 & Extension of Balakrishnan \ac{PDF} perturbation formula & 4 & No & No & No \\\addlinespace[4pt] \hline\addlinespace[4pt]
         Jones skew t
         \cite{jones2003skew} & 2003 & Two-piece construction of differing tail kurtosis (df ) & 4 & Partially & Partially & Yes \\\addlinespace[4pt] \hline\addlinespace[4pt]
         Skew-generalised normal
         \cite{arellano2004} & 2004 & Extension of Azzalini \ac{PDF} perturbation formula & 5 & No & No & No \\\addlinespace[4pt] \hline\addlinespace[4pt]
         Flexible class of skew symmetric distributions\cite{MaGenton2004}
          & 2004 & Extension of Azzalini \ac{PDF} perturbation with k’th odd order polynomial & $k\in\{1, 2, 3, 5\dots\}$& No & No & No \\\addlinespace[4pt] \hline\addlinespace[4pt]
         Generalised Balakrishnan skew normal
         \cite{gupta2004generalized} & 2004 & Extension of Balakrishnan \ac{PDF} perturbation formula & 5 & No & No & No\\\addlinespace[4pt] \hline\addlinespace[4pt]
         Generalised skew distributions\cite{huang2007generalized} & 2007 & Transformation with constraints  & Not derived  & No  & Possibly & No \\\addlinespace[4pt] \hline\addlinespace[4pt]
         Sinh-arcsinh \cite{jones2009sinh}& 2009 & Transformation of variable & 4 & Partially & Yes & Yes  \\\addlinespace[4pt] \hline\addlinespace[4pt]
         Two-parameter Balakrishnan skew-normal \cite{bahrami2009two} & 2009 & Extension of Balakrishnan \ac{PDF} perturbation formula & 6 & No & No & No \\\addlinespace[4pt] \hline\addlinespace[4pt]
         Alpha skew normal
         \cite{elal2010alpha} & 2010 & Perturbation of \ac{PDF} & 4 & No & No & No\\\addlinespace[4pt] \hline\addlinespace[4pt]
         Extended skew generalised Normal Distribution \cite{venegas2011extension} & 2011 & Extended Skew Azzalini Normal Distribution & 5 & No & No & No \\\addlinespace[4pt] \hline\addlinespace[4pt]
         Beta generated \cite{alexcord11} &  & Generalisation of order statistical scaling & 4 & No & Yes & Possibly \\\addlinespace[4pt] \hline\addlinespace[4pt]
         Flexible skew-generalised normal\cite{nekoukhou2013flexible} & 2013 & Extension of Azzalini \ac{PDF} perturbation with k’thodd order polynomial & $k\in\{1, 2, 3, 5\dots\}$ & No & No & Possibly\\\addlinespace[4pt] \hline\addlinespace[4pt]
%    \end{tabular}
\end{tabularx}
\label{tab:skewmech}
\end{table}
The recent and innovative \ac{BTN} \cite{BTGN} extension of the \ac{EP} distribution makes it possible to skew the \ac{EP} by asymmetric kurtosis without a "joint", enabling lighter or heavier than baseline tails, finite moments, interpretable parameters, closed form \ac{PDF}, trouble-free estimation, and parsimony. This is due to the \ac{BTN} 's parameters that specifically control body and tail shape. The derivative kernel skewing mechanism uses the derivative kernel paradigm employed to derive the \ac{BTN}  and results in a method that is a combination of the two-piece approaches, skewing with kurtosis while making an appropriate adjustment in scale.

The chapter is structured as follows. Section \ref{sec:skewmot} introduces the skewing mechanism for the \ac{FIN} distribution using the derivative kernel paradigm. Section \ref{sec:fin} provides derivations of statistical properties such as cumulative probability function (\ac{CDF}), moments, and \ac{ML} estimation. Section \ref{sec:application} applies the \ac{FIN} in a multivariate context to stock returns data using copulas.

\section{Derivative kernel skewing method}
The \ac{FIN} kernel is based on the ``derivative kernel” paradigm developed by \cite[pg. 3]{BTGN}:
\vspace{3mm}\\
\say{\textit{The construction is based on the relationship between the derivative kernel function and the \ac{PDF}. Given some ``appropriate'' derivative kernel function, $k'(z)$,  a new distribution can be generated by simply integrating $k'(z)$ and normalising the resulting function to give a new \ac{PDF} $f(z)$.}
}
\vspace{3mm}\\
The process of generating a new distribution involves four steps:
\begin{enumerate}
    \item Investigate the functional properties of derivative kernel functions from existing distributions.
    \item Construct a derivative kernel function with the desired functional form.
    \item Take the indefinite integral of the derivative kernel function to obtain a new kernel function.
    \item Normalise the kernel function to obtain a new \ac{PDF} and hence a distribution.
\end{enumerate}
As this was a quite complex task, both the derivative kernel function and the kernel function were considered concurrently during the specification of the \ac{FIN} distribution. Steps 1 to 3 can be found in this section, with step 4 in Section \ref{sec:fin}.
\subsection*{Step 1}
The \ac{PDF} and derivative \ac{PDF} functions for the \ac{BTN} , two-piece normal, and Azzalini skew normal distributions are investigated here. The sinh-arcsinh and two-piece kurtosis skewing of the Jones skew t have the most desirable skewing mechanisms, as shown in Table \ref{tab:skewmech}. To improve upon these, it is essential to retain the interpretation of the shape body parameter, allow for lighter than normal tails, and ensure finite moments for the domain of the shape parameters. An examination of the \ac{BTN}  distribution will demonstrate how these improvements can be achieved. The remaining distributions were chosen as they are the best-performing skewing mechanisms for the \ac{EP} in Table \ref{tab:PNskewtimeline}. When considered together, these three studies form the basis for the construction of the \ac{FIN} derivative kernel. Since the shape of the \ac{PDF} of a distribution is mathematically proportional to the shape of its kernel function, either can be used to gain insights into the functional properties of the kernel function.

\begin{figure}[ht!]
    \centering
    \includegraphics[width=\linewidth,keepaspectratio]{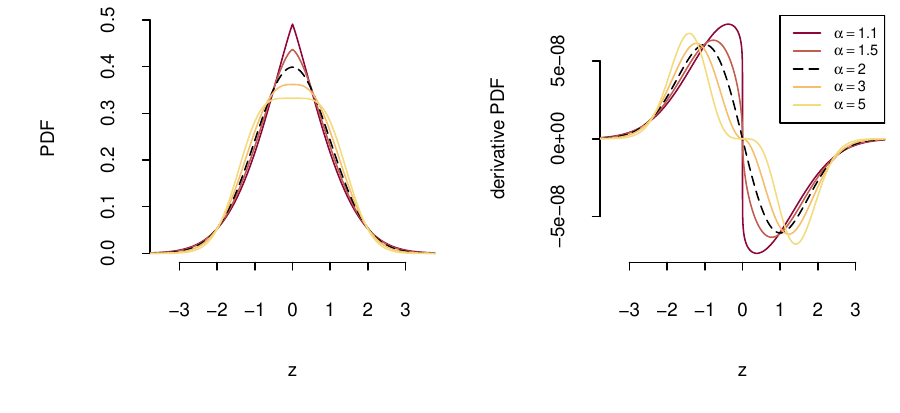}
    \caption{The \ac{PDF} and derivative \ac{PDF} of the \ac{BTN}  distribution for a fixed normal tail shape $\beta=2$ and various body shape values of $\alpha$.}
    \label{fig:btnbody}
\end{figure}

\begin{figure}[ht!]
	\centering{
		\includegraphics[width=12cm,keepaspectratio]{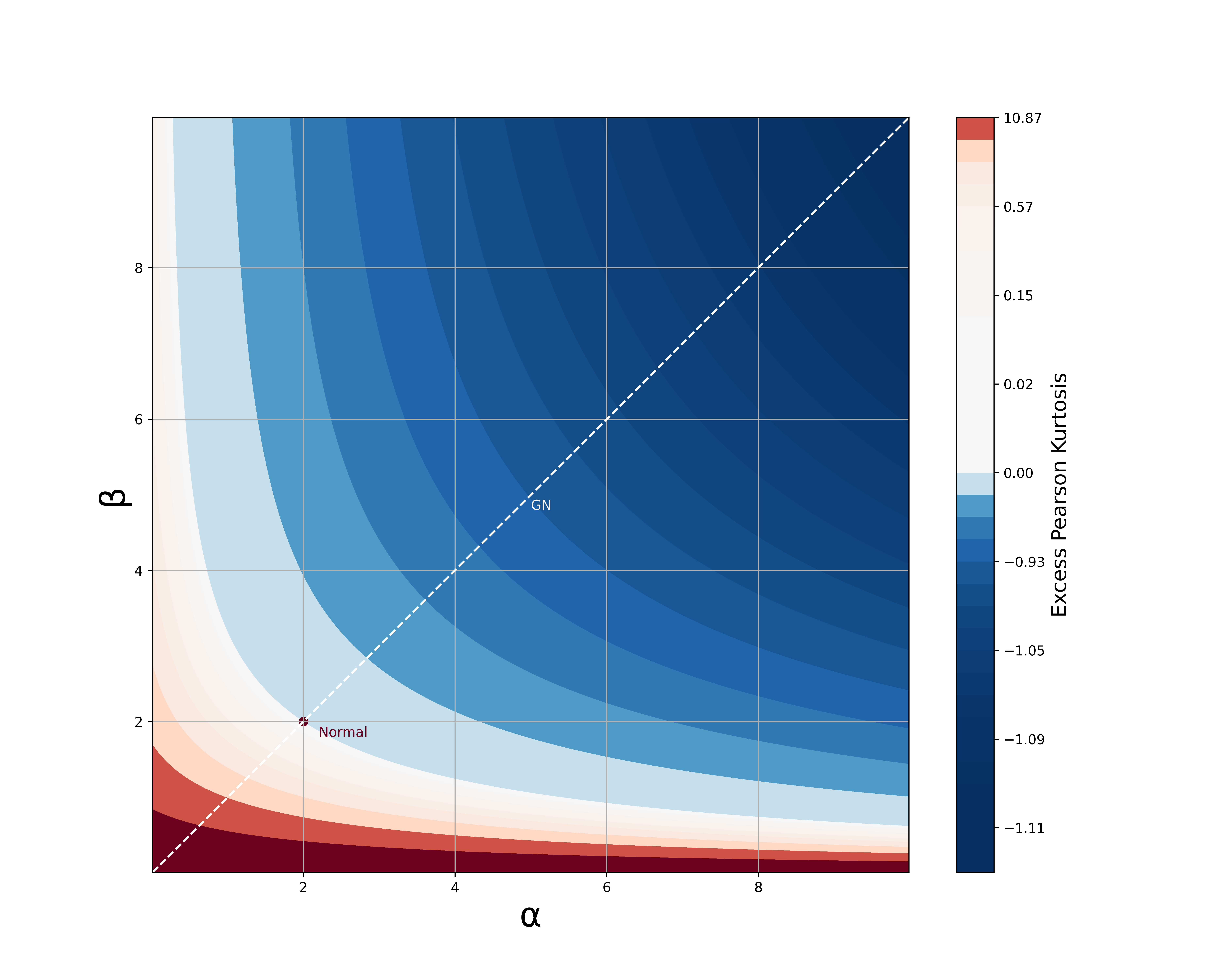}
	}
	\caption{ The excess Pearson kurtosis of the \ac{BTN} distribution for different values of $\alpha$ and $\beta$. Excess kurtosis is equal to zero, when it is equal to that of the normal distribution.}
	\label{fig:kurtosisheatmap}
\end{figure}

\begin{figure}
    \centering
    \includegraphics[width=\linewidth,keepaspectratio]{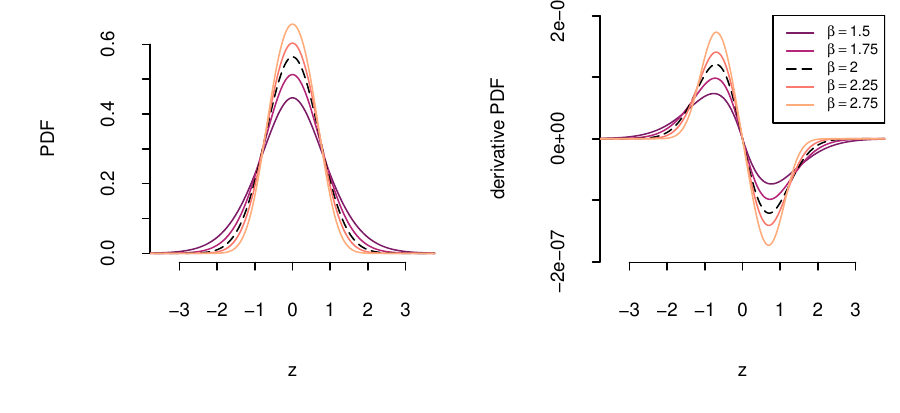}
    \caption{The \ac{PDF} and derivative \ac{PDF} of the \ac{BTN}  distribution for a fixed normal body shape $\alpha=2$ and various tails shape values of $\beta$.}
    \label{fig:btntail}
\end{figure}

\begin{figure}[h!]
    \centering
    \includegraphics[width=\linewidth,keepaspectratio]{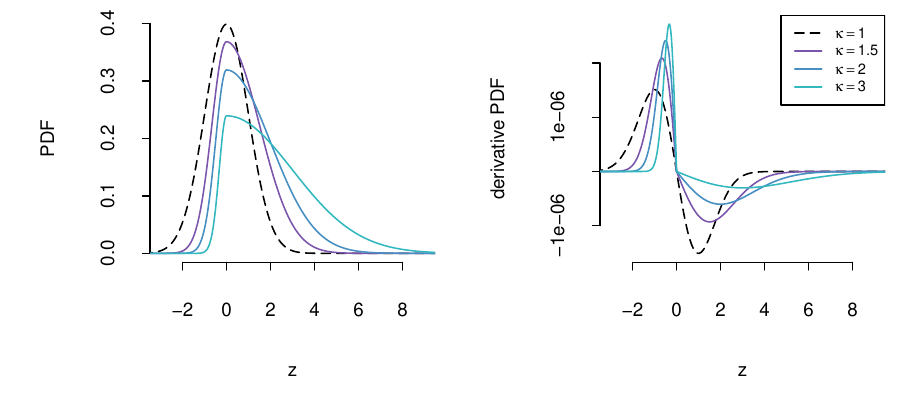}
    \caption{The \ac{PDF} and derivative \ac{PDF} of the two-piece normal distribution for varying values of the skewness parameter $\kappa$. The dashed line represents the symmetric case.}
    \label{fig:tpnorm}
\end{figure}

\begin{figure}[h!]
    \centering
    \includegraphics[width=\linewidth,keepaspectratio]{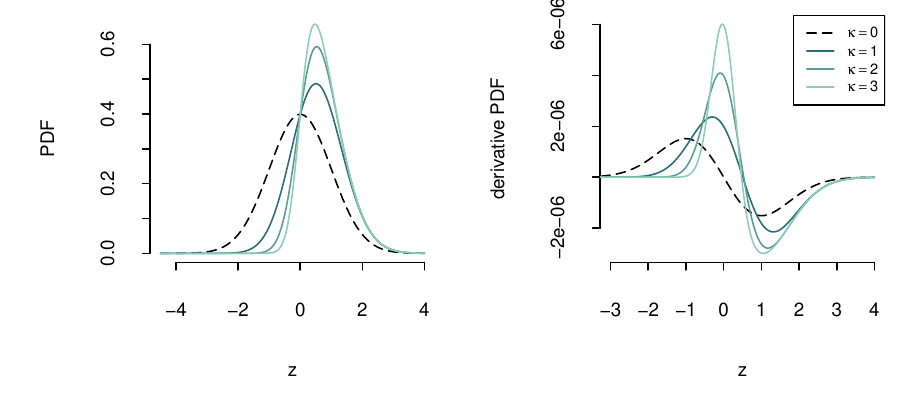}
    \caption{The \ac{PDF} and derivative \ac{PDF} of the Azzalini skew normal distribution for varying values of the skewness parameter $\kappa$. The dashed line represents the symmetric case.}
    \label{fig:azzaskewnorm}
\end{figure}

The first distribution we investigate is the distribution we wish to generalise the \ac{BTN}. The \ac{BTN}  derivative kernel function is given by
\begin{equation}
 k'(z;\alpha,\beta)=-\beta\text{ sign}(z) |z|^{\alpha-1}\text{e}^{-|z|^\beta},
 \label{eq:btnderker}
\end{equation}
the kernel function is given by
\begin{equation}
k(z;\alpha,\beta)= \int k'(z)dz
  =\int- \beta z^{\alpha-1}\text{e}^{-z^\beta} dz
  =\Gamma\left(\frac{\alpha}{\beta},z^\beta \right),
 \label{eq:btnker} \notag
\end{equation} 
where $\Gamma\left(\cdot,\cdot \right)$ is the upper incomplete gamma function, see \cite[pg. 899]{Gr<3dstheyn}.
\\The \ac{PDF} is given by
\begin{equation}
 f(z;\alpha,\beta)=\frac{\Gamma\left(\frac{\alpha}{\beta},|z|^\beta \right)}{2\Gamma\left( \frac{\alpha+1}{\beta}\right)},
 \label{eq:btpdf} \notag
\end{equation}
where $z\in\mathbb{R}$, $\alpha, \beta>0$. The parameter $\alpha$ has the interpretation of body shape, and the parameter $\beta$ has the interpretation of tail shape. The \ac{EP} distribution is nested within the \ac{BTN} for values of $\alpha=\beta$, and the normal distribution is given when $\alpha=\beta=2$. In Figure \ref{fig:btnbody} the \ac{BTN} is depicted with a fixed tail shape and varying ``sharper" and ``flatter" body shapes.
Figure \ref{fig:kurtosisheatmap} best illustrates what the \ac{BTN} achieves in terms of a body and tail generalisation of the \ac{EP} distribution. Although different levels of kurtosis can be achieved along straight line combinations of $\alpha=\beta$, i.e. the \ac{EP} distribution, where the body and tail shapes cannot be varied. The range of body and tail shapes for a given level of kurtosis is a contour on the plane of $\alpha,\beta>0$.
Figure \ref{fig:btnbody} additionally shows the relationship between different distribution body shapes and their derivative \ac{PDF} for the \ac{BTN} \ac{PDF}. The authors of \cite{BTGN} determined that the shape of a distribution's body is determined by the first factor $|z|^{\alpha-1}$ in (\ref{eq:btnderker}) that coincides with the approximate region of $-1.5<z<1.5$ in Figure \ref{fig:btnbody}. The derivative \ac{PDF} magnitudes are higher and lower in the region around zero, the body of the distribution. This is reflected in the "sharper" and "flatter" body shapes. Figure \ref{fig:btntail} shows the relationship between different distribution tail shapes and their derivative \ac{PDF} for the \ac{BTN}  \ac{PDF}.
The authors of \cite{BTGN} determined that the shape of a distribution's tail is determined by the second factor $\text{e}^{-|z|^\beta}$ in (\ref{eq:btnderker}). The derivative \ac{PDF} magnitudes are higher and lower in the tail region for "heavier" and "lighter" tail shapes, respectively. The constant factor $-\beta\text{ sign}(z)$ in (\ref{eq:btnderker}) is also important, as its disturbance can lead to the "joint" at the mode of the distribution, as is seen in the two-piece normal distribution shown in Figure \ref{fig:tpnorm}. Thus, for the \ac{BTN}  kernel function, the factors containing $z$ as arguments are the determinants of the shape of the distribution.
Examining the two-piece normal, see Table \ref{tab:skewmech}, derivative kernel function elucidates why this happens. The derivative kernel function for the two-piece normal distribution is given by 
\begin{equation}
k(z;\kappa)= \left\{\begin{array}{ll}
    2\kappa^{2} |z|\text{e}^{-(\kappa z)^2}
        & z \leq 0 \\
    -2\frac{1}{\kappa^2} z\text{e}^{-(z/\kappa)^2}
        & z > 0 \\
    \end{array}
    \right.,
    \label{eq:tpnormderkernel}
\end{equation}
where $z\in\mathbb{R}$, and $\kappa>0$ is the skewness parameter. Near the origin, the $e^{-x}$ function has a slope of negative one, as defined by the natural exponential function.  Furthermore noting the following factor present in (\ref{eq:tpnormderkernel}), 
\begin{equation}
f(z)=
    \left\{\begin{array}{ll}
    2|z|
        & z \leq 0 \\
    -2z
        & z > 0 \\
    \end{array}
    \right.
=-2z,\forall z\in\mathbb{R}, \notag
\end{equation}
we can observe that the factor $-2z$ is the same for both $z\leq0$ and $z>0$.
The origin of the ``joint" is traced to the different constant factors $\kappa^2$ on the \ac{LHS} of (\ref{eq:tpnormderkernel}) and $\frac{1}{\kappa^2}$ on the \ac{RHS} of zero in (\ref{eq:tpnormderkernel}). Note that this ``joint" is even more pronounced in the two-piece shape skewing versions of \cite{zhu2009properties,bottazzi11} due to the added effect of different body shapes. The \ac{BTN}  kernel with asymmetric tail weights is a possible solution to the problem, as it allows the body shape of the distribution to remain the same on both sides. However, one must take care to adhere to the functional form of (\ref{eq:btnderker}) strictly, as any deviation from it could lead to similar issues. Finally, we consider the Azzalini skew normal, see Table \ref{tab:skewmech}, \ac{PDF}, and derivative \ac{PDF} in Figure \ref{fig:azzaskewnorm}. This Azzalini skewing method has a smooth shape without any ``joint" at zero. However, the mode does not coincide with the location parameter, and the deviation increases with higher amounts of skewness. In terms of the derivative kernel \ac{PDF}, the skewness originates from an asymmetrical kurtosis. This is visible in the fact that for increasing skewness of the distribution, lower derivative \ac{PDF} magnitudes on one side are accompanied by higher values on the other side. This confirms that kurtosis skewing is a suitable way of skewing, provided the body shape of the distribution can be maintained.
\subsection*{Steps 2 and 3}
\label{sec:skewmot}
\begin{figure}[h!]
    \centering
    \includegraphics[width=\linewidth,keepaspectratio]{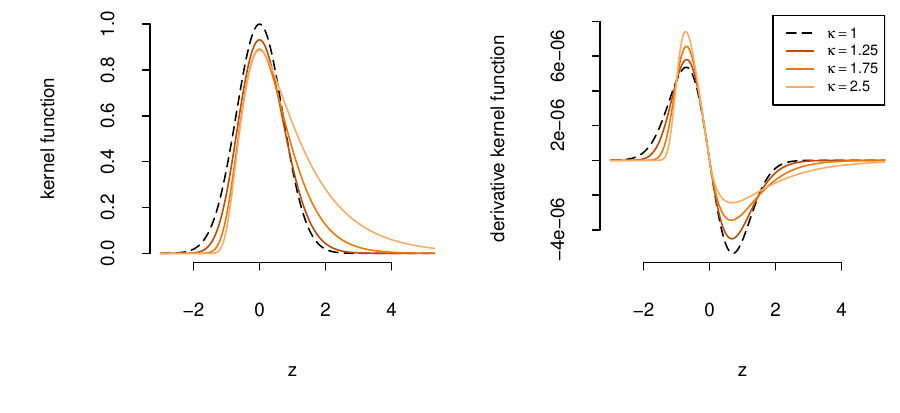}
    \caption{The kernel and derivative kernel of the \ac{FIN} distribution for varying values of the skewness parameter $\kappa$. The dashed line represents the symmetric case.}
    \label{fig:finkern}
\end{figure}

The main elements of the \ac{FIN} kernel function are an asymmetric adjustment of the left and right tail shapes using $\kappa$, a continuity adjustment $\phi$, and a scaling adjustment for the right tail factor, $\phi^{-\frac{1}{\alpha}}$ due to differing kurtosis. These components were determined by mathematical inspection iterating between adding a component and evaluating the effect on the kernel and derivative kernel function. Each component's latter function is given is subsequently proven. The kernel of the \ac{FIN} is defined as
\begin{equation}
    k(z;\alpha,\beta,\kappa) = \left\{\begin{array}{ll}
    \kappa^{-1} \Gamma \left( \frac{\alpha}{\beta \kappa}, |z|^{\beta \kappa} \right)
        & z \leq 0 \\
    \phi \kappa \Gamma \left( \frac{\alpha \kappa}{\beta}, \left( \phi^{-\frac{1}{\alpha}} z \right)^{\beta / \kappa} \right)
        & z > 0 \\
    \end{array}
    \right.,\label{eq:FINkern}
\end{equation}
where $\phi=\frac{\Gamma \left ( \frac{\alpha}{\beta \kappa}\right )}{\Gamma \left ( \frac{\alpha \kappa}{\beta}\right )\kappa^2}$ and $\alpha,\beta,\kappa > 0$.
As $\kappa$ increases, the left tail coefficient $\beta\kappa$ increases and decreases the left tail weight of (\ref{eq:FINkern}). Conversely, an increase in $\kappa$ decreases the right tail coefficient $\beta/\kappa$, increases the right tail weight. Therefore, an increase in $\kappa$, the skewness parameter, increases the positive skewness of the distribution. The continuity adjustment $\phi$ ensures that the kernel is continuous at the mode. Evaluating the following limit
\begin{equation}
\lim_{z\to0^+} \phi \kappa \Gamma \left( \frac{\alpha \kappa}{\beta}, \left( \phi^{-\frac{1}{\alpha}} z \right)^{\beta / \kappa} \right)
=\frac{\Gamma \left ( \frac{\alpha}{\beta \kappa}\right )}{\Gamma \left ( \frac{\alpha \kappa}{\beta}\right )\kappa^2}\kappa\Gamma \left ( \frac{\alpha \kappa}{\beta}\right )
=\kappa^{-1}\Gamma \left ( \frac{\alpha \kappa}{\beta}\right )
=k(0;\alpha,\beta,\kappa), \notag
\end{equation}
confirms the continuity at the mode. The scaling correction for the \ac{RHS} of $\phi^{-\frac{1}{\alpha}}$ within $\Gamma(\cdot,\cdot)$ ensures that the two halves of the distribution are using the same scale. This is accomplished by ensuring that the derivative of the kernel function has the same functional form as the derivative kernel function of the \ac{BTN}. Demonstrating this, the derivative of the kernel function (\ref{eq:FINkern}) is given by
\begin{align}
    \frac{\partial}{\partial z} k(z;\alpha,\beta,\kappa) 
    &= \left\{\begin{array}{ll}
    \frac{\partial}{\partial z}\left(\kappa^{-1} \Gamma \left( \frac{\alpha}{\beta \kappa}, |z|^{\beta \kappa} \right)\right)
        & z \leq 0 \\
    \frac{\partial}{\partial z}\left(\phi \kappa \Gamma \left( \frac{\alpha \kappa}{\beta}, \left( \phi^{-\frac{1}{\alpha}} z \right)^{\beta / \kappa} \right)\right)
        & z > 0 \notag \\
    \end{array}
    \right., \\
    &= \left\{\begin{array}{ll}
    \kappa^{-1} (-z^{\beta\kappa(\frac{\alpha}{\beta\kappa}-1)}e^{-|z|^{\beta\kappa}})(-\beta\kappa z^{\beta\kappa-1})
        & z \leq 0 \\
    \phi \kappa \left(-\left(\phi^{-\frac{1}{\alpha}} z \right)^{\frac{\beta}{\kappa}(\frac{\alpha\kappa}{\beta}-1)} e^{-\left(\phi^{-\frac{1}{\alpha}} z \right)^{\beta/\kappa}}\frac{\beta}{\kappa}\left(\phi^{-\frac{1}{\alpha}} z \right)^{\frac{\beta}{\kappa}-1}\phi^{-\frac{1}{\alpha}}\right)
        & z > 0 \notag \\
    \end{array}
    \right.,\\
    &= \left\{\begin{array}{ll}
    \kappa^{-1} \left(-z^{(\alpha-\beta\kappa+\beta\kappa-1)}e^{-|z|^{\beta\kappa}}\beta\kappa\right)
        & z \leq 0 \\
    \phi \kappa \left(-\phi^{(-1+\frac{\beta}{\alpha\kappa}-\frac{\beta}{\alpha\kappa}+\frac{1}{\alpha}-\frac{1}{\alpha})} z ^{(\alpha-\frac{\beta}{\kappa}+\frac{\beta}{\kappa}-1)} e^{-\left(\phi^{-\frac{1}{\alpha}} z \right)^{\beta/\kappa}}\frac{\beta}{\kappa}\right)
        & z > 0  \notag \\
    \end{array}
    \right., \\
    &= \left\{\begin{array}{ll}
    \beta z^{\alpha-1}e^{-|z|^{\beta\kappa}}
        & z \leq 0 \\
    -\beta z ^{\alpha - 1} e^{-\left(\phi^{-\frac{1}{\alpha}} z \right)^{\beta/\kappa}}
        & z > 0 \\
    \end{array}
    \right. .
    \label{eq:FINderkern}
\end{align}
Both sides of (\ref{eq:FINderkern}) contain the same factors, namely $\beta$, $z^{\alpha-1}$, and $\text{e}^{-f(z)}$, which are also present in the \ac{BTN}  derivative kernel (\ref{eq:btnderker}). The \ac{FIN} kernel has a single maximum value at $z=0$, meaning no additional re-parametrisation is necessary to interpret the roles of the parameters. This is further demonstrated in Figure \ref{fig:finkern}.
The \ac{FIN} kernel has a smooth shape with no ''joints" and a distribution mode equal to its location parameter. The skewness parameter has little effect on the distribution's body shape, which is virtually identical to the symmetric distribution. Properties carried over from the \ac{BTN}  distribution include interpretable parameters, finite moments, and tractable equations  \cite{BTGN}.
\section{The flexible interpretable normal distribution}
\label{sec:fin}
In this section, the \ac{FIN} is defined and various properties are derived, such as the
\ac{PDF}, \ac{CDF}, moments, and \ac{ML} equations.
\subsection{PDF}
\begin{figure}[ht!]
	\centering{
		\includegraphics[width=\linewidth,keepaspectratio]{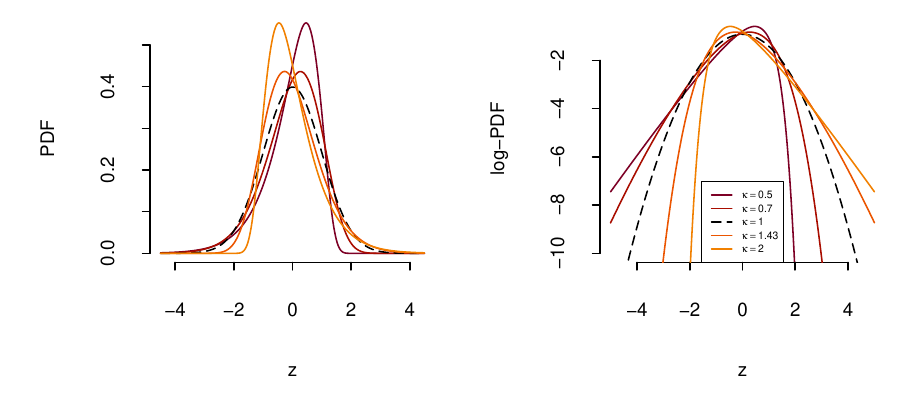}
	}
\caption{The  \ac{PDF} of the \ac{FIN} distribution for fixed normal distribution body and tail shapes $\alpha=\beta=2$, and varying values of $\kappa$. The mean and variance of each \ac{PDF} are set equal to zero and one, respectively.}
	\label{fig:finkapp}
\end{figure}

\begin{figure}[ht!]
	\centering{
		\includegraphics[width=\linewidth,keepaspectratio]{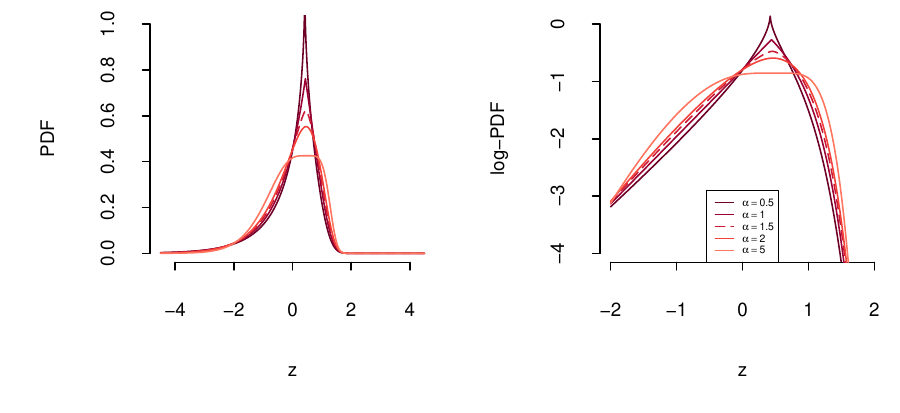}
	}
\caption{The \ac{PDF} of the \ac{FIN} distribution for fixed tail shape $\beta=2$ and skewness parameter of $\kappa=0.9$, and varying values of body shape $\alpha$. The mean and variance of each \ac{PDF} are set equal to zero and one, respectively.}
\label{fig:finalpha}
\end{figure}

\begin{figure}[ht!]
	\centering
	\includegraphics[width=\linewidth,keepaspectratio]{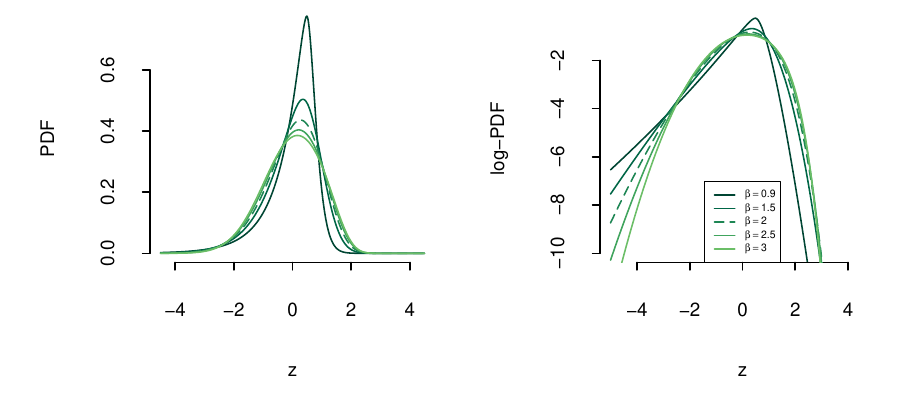}
	\caption{The PDF of the FIN distribution for a fixed body shape $\alpha=2$ and skewness parameter of $\kappa=0.9$, and varying values of tail shape $\beta$. The mean and variance of each PDF are set equal to zero and one, respectively.}
	\label{fig:finbeta}
\end{figure}

Let the standard\footnote[5]{Location and scale values are equal to zero and one, respectively.}  \ac{FIN} kernel function be $k(\cdot)$ be defined by (\ref{eq:FINkern}). The normalisation constant $\delta$ for $k(\cdot)$ is obtained by direct substitution of %(\ref{eq:lem3}) 
Lemma 3, in the Appendix, for the integral
\begin{equation}
\delta\stackrel{set}{=}
    \int^{\infty}_{-\infty}{k(z) dz}=
    \kappa^{-1}\Gamma \left( \frac{\alpha + 1}{\beta \kappa} \right) 
    + \phi^{\frac{\alpha + 1}{\alpha}}\kappa
    \Gamma \left( \frac{\alpha + 1}{\beta}\kappa \right)\notag. 
\end{equation}
The \ac{PDF} of the standard \ac{FIN}, denoted by $Z\sim FIN(\alpha, \beta, \kappa)$, is given by 
\begin{align}
    f(z;\alpha,\beta,\kappa)&= \left\{\begin{array}{ll}
    \frac{\kappa^{-1} }{ \delta }\Gamma \left( \frac{\alpha}{\beta \kappa}, |z|^{\beta \kappa} \right)
        & z \leq 0 \\
    \frac{\phi \kappa}{ \delta } \Gamma \left( \frac{\alpha \kappa}{\beta}, \left( \phi^{-\frac{1}{\alpha}} z \right)^{\beta / \kappa} \right)
        & z > 0 \\
    \end{array}
    \right.,\label{eq:stanFIN} 
\end{align}
where $z\in\mathbb{R}$, $\alpha, \beta, \kappa>0$.
From the transformation $X = \mu+\sigma Z$, the non-standard\footnote[6]{Arbitrary location and scale values.} \ac{FIN} \ac{PDF}, denoted by $X\sim FIN(\mu, \sigma, \alpha, \beta,\kappa)$, is given by 
\begin{equation}
    f(x;\mu,\sigma,\alpha,\beta,\kappa)= \left\{\begin{array}{ll}
    \frac{\kappa^{-1} }{ \delta \sigma}\Gamma \left( \frac{\alpha}{\beta \kappa}, |z|^{\beta \kappa} \right)
        & z \leq 0 \\
    \frac{\phi \kappa}{ \delta \sigma} \Gamma \left( \frac{\alpha \kappa}{\beta}, \left( \phi^{-\frac{1}{\alpha}} z \right)^{\beta / \kappa} \right)
        & z > 0 \\
    \end{array}
    \right.,\label{eq:lsFIN}
\end{equation}
where $x\in\mathbb{R}$, $z=\frac{x-\mu}{\sigma}$, $\sigma, \alpha, \beta, \kappa>0$.
%
%\subsubsection{Parameter roles}
In Figure \ref{fig:finkapp}, the role of the new skewness parameter $\kappa$ on the \ac{FIN} \ac{PDF} is illustrated. For values of $\kappa<1$, the distribution is skewed towards to the left, when $\kappa=1$ the distribution is symmetric, and for $\kappa>1$ the distribution is skewed towards the right.
Next, we verify that the role of the body and tail shape parameters $\alpha$ and $\beta$ have been preserved from the \ac{BTN}. In Figure \ref{fig:finalpha}, the \ac{FIN} \ac{PDF} is depicted for varying values of $\alpha$. For values of $\alpha<2$ the \ac{FIN} has a ``sharper" than normal body shape, when $\alpha=2$ the body shape is normal, and for $\alpha>2$ the body shape is ``flatter" than normal.
In Figure \ref{fig:finbeta}, the \ac{FIN} \ac{PDF} is depicted for varying values of $\beta$. For increasing values of $\beta$ the \ac{FIN} distribution has lighter tails overall and heavier tails for decreasing values of $\beta$. If the tails of the \ac{FIN} are to be compared to the normal distribution, the left tail coefficient $\beta\kappa$, and right tail coefficient $\beta/\kappa$  of the \ac{FIN} need to be interpreted separately. For a tail coefficient of less than two, the specific tail has a heavier than normal tail weight, for a coefficient equal to two, the tail weight is normal, and for a coefficient greater than two, the tail weight is lighter than normal.

%

%\subsubsection{submodels}
Focusing on the submodels of the \ac{FIN}, several kurtosis skewed versions of popular distributions such as the normal, Laplace, and PN were created during the process of developing the \ac{FIN} distribution. In Table \ref{tab:finsubmods}, the symmetric submodels are given where $\kappa=1$. Each symmetric distribution has a kurtosis skewed counterpart simply by varying the skewness parameter by $\kappa\neq 1$. 
\begin{table}[ht!]
\centering
\begin{tabular}{llllll}
\hline
Distribution                 & $\mu$ & $\sigma$     & $\alpha$ & $\beta$ & $\kappa$ \\
\hline
Normal                       & $\mu$ & $\sigma/\sqrt{2}$ & 2        & 2       & 1        \\
Laplace                      & $\mu$ & $\sigma$     & 1        & 1       & 1        \\
Power normal                 & $\mu$ & $\sigma$     & $s$      & $s$     & 1        \\
Flexible tail normal         & $\mu$ & $\sigma/\sqrt{2}$ & 2        & $\beta$ & 1        \\
Body-tail generalised normal & $\mu$ & $\sigma$     & $\alpha$ & $\beta$ & 1       
\end{tabular}
\label{tab:finsubmods}
\caption{Nested symmetrical models of the \ac{FIN} distribution.}
\end{table}\\
The \ac{FTN} distribution is an important submodel that is obtained by setting the body shape parameter $\alpha=2$ in the \ac{FIN} distribution. This distribution has four parameters that characterise its shape: location, scale, tail weight, skewness, and a fixed normal body shape. In situations where the modelling task is less demanding or a simpler distribution is preferred, the \ac{FTN} would be appropriate. The lighter-than-normal tail property and finite moments make it a strong alternative to the t distribution. The \ac{PDF} of the non-standard \ac{FTN} \ac{PDF} is given by 
\begin{equation}
    f(x;\mu,\sigma,\beta,\kappa)= \left\{\begin{array}{ll}
    \frac{\kappa^{-1} }{ \delta \sigma}\Gamma \left( \frac{2}{\beta \kappa}, |z|^{\beta \kappa} \right)
        & z \leq 0 \\
    \frac{\phi \kappa}{ \delta \sigma} \Gamma \left( \frac{2 \kappa}{\beta}, \left( \phi^{-\frac{1}{2}} z \right)^{\beta / \kappa} \right)
        & z > 0 \\
    \end{array}
    \right.,\label{eq:lsFTN}\notag
\end{equation}
where $x\in\mathbb{R}$, $z=\frac{x-\mu}{\sigma}$, $\sigma, \beta, \kappa>0$, $\phi=\frac{\Gamma \left ( \frac{2}{\beta \kappa}\right )}{\Gamma \left ( \frac{2 \kappa}{\beta}\right )}\kappa^{-2}$, $\delta =  \kappa^{-1} \Gamma \left( \frac{ 3}{\beta \kappa} \right) + \phi^{\frac{ 3}{2}} \kappa \Gamma \left( \frac{ 3}{\beta}\kappa \right)$. The remaining statistical quantities for the \ac{FTN} can be determined by substituting $\alpha=2$ into the respective equations of the \ac{FIN} distribution.
\subsection{Cumulative distribution function}
The \ac{CDF} of the standard \ac{FIN} is derived with the definition of a \ac{CDF} and (\ref{eq:stanFIN}) working in a piece-wise fashion. For $z \leq 0$ substituting %(\ref{eq:lem2})
Lemma 2, in the Appendix, we have that:
\begin{align}
F(z;\alpha,\beta,\kappa)
%&=\int^{z}_{-\infty}{f(t) dt} \notag \\ 
        &=\int^{z}_{-\infty}{\frac{ \kappa^{-1}}{\delta} \Gamma \left ( \frac{\alpha}{\beta \kappa}, |t|^{\beta\kappa} \right ) dt } \notag\\
        &={\frac{ \kappa^{-1}}{\delta}\int^{\infty}_{z} \Gamma \left ( \frac{\alpha}{\beta \kappa}, t^{\beta\kappa} \right ) dt }\notag\\
        &=\frac{\kappa^{-1}}{\delta} \left ( \Gamma \left ( \frac{\alpha +1}{\beta \kappa}, z ^{\beta\kappa} \right )
        - z \Gamma \left ( \frac{\alpha}{\beta \kappa} , z^{\beta\kappa} \right )
        \right ).\label{eq:cdflhs}
    \end{align}
For $z>0$:
\begin{equation}
F(z;\alpha,\beta,\kappa)%1-\int^{\infty}_{z}{f(t) dt }
=1-\int^{\infty}_{z}{\phi \kappa \Gamma \left( \frac{\alpha \kappa}{\beta}, \left( \phi^{-\frac{1}{\alpha}}z \right)^{\beta/\kappa} \right) dt}.
\label{eq:cdfrhs1}
\end{equation}
%Letting $s=\phi^{-\frac{1}{\alpha}}t$, that implies, $t=\phi^{\frac{1}{\alpha}}s$ and $\frac{\partial}{\partial s}\phi^{\frac{1}{\alpha}}s=\phi^{\frac{1}{\alpha}}$.
%Letting $s=\phi^{-\frac{1}{\alpha}}t$, that implies, $t=\phi^{\frac{1}{\alpha}}s$, and substituting %(\ref{eq:lem2})
%Lemma 2, in the Appendix, the integral in (\ref{eq:cdfrhs1}) becomes:
Letting $s=\phi^{-\frac{1}{\alpha}}t$, and substituting %(\ref{eq:lem2})
Lemma 2, in the Appendix, the integral in (\ref{eq:cdfrhs1}) becomes:
\begin{align}
    &=1-\phi \kappa\int^{\infty}_{\phi^{-\frac{1}{\alpha}}z}{\phi^{\frac{1}{\alpha}}\Gamma \left( \frac{\alpha\kappa}{\beta}, s^\frac{\beta}{\kappa} \right) ds }\notag\\
    &=
    1-\phi^{\frac{\alpha+1}{\alpha}} \left (
    \Gamma\left( \frac{\alpha+1}{\beta/\kappa},\left(\phi^{-\frac{1}{\alpha}}z\right)^{\beta/\kappa}\right)-\left (\phi^{-\frac{1}{\alpha}}z \right)\Gamma\left( \frac{\alpha}{\beta/\kappa},\left(\phi^{-\frac{1}{\alpha}}z\right)^{\beta/\kappa}\right)
    \right ).\label{eq:cdfrhs}
\end{align}
Summarising from (\ref{eq:cdflhs}) and (\ref{eq:cdfrhs}) we have that the \ac{CDF} of the standard \ac{FIN} distribution is given by:
\begin{equation}
    F(z;\alpha,\beta,\kappa) = \left\{\begin{array}{ll}
    \frac{\kappa^{-1}}{\delta} \left ( \Gamma \left ( \frac{\alpha +1}{\beta \kappa}, z ^{\beta\kappa} \right )
    -z \Gamma \left ( \frac{\alpha}{\beta \kappa} , z^{\beta\kappa} \right )
    \right )
    & z \leq 0 \\
    1-\phi^{\frac{\alpha+1}{\alpha}} \left (
    \Gamma\left( \frac{\alpha+1}{\beta}\kappa,\left(\phi^{-\frac{1}{\alpha}}z\right)^{\beta/\kappa}\right)-\left (\phi^{-\frac{1}{\alpha}}z \right)\Gamma\left( \frac{\alpha\kappa}{\beta},\left(\phi^{-\frac{1}{\alpha}}z\right)^{\beta/\kappa}\right)
    \right )
    & z > 0 \\
    \end{array}
    \right. \label{eq:FINcdf}.
    \end{equation}
 Subsequently, the \ac{CDF} of $X\sim \ac{FIN}(\mu,\sigma,\alpha,\beta,\kappa)$ is given by the substitution of $z=\frac{x-\mu}{\sigma}$ in (\ref{eq:FINcdf}).
 \subsection{Moments}
The $r$-th moment of the standard \ac{FIN} is derived from (\ref{eq:FINkern}), and substituting %(\ref{eq:lem4})
Lemma 4, in the Appendix, we have that:
\begin{align}
E(Z^r)&=
    \int^{\infty}_{-\infty}{z^r\frac{k(z)}{\delta} dz}\notag\\
&=\frac{(-1)^r\frac{ \kappa^{-1}}{r+1} \Gamma \left( \frac{r + \alpha + 1}{\beta \kappa} \right) 
    + \phi^{\frac{r + \alpha + 1}{\alpha}}\frac{\kappa}{r+1} 
    \Gamma \left( \frac{r + \alpha + 1}{\beta}\kappa \right)}
    { \kappa^{-1} \Gamma \left( \frac{ \alpha + 1}{\beta \kappa} \right) + \phi^{\frac{ \alpha + 1}{\alpha}} \kappa \Gamma \left( \frac{ \alpha + 1}{\beta}\kappa \right)}.\label{eq:mom}
\end{align}
Subsequently, the $r$th moment $X\sim \ac{FIN}(\mu,\sigma,\alpha,\beta,\kappa)$ is a function of (\ref{eq:mom}) and the binomial expansion of a polynomial (\cite{Gr<3dstheyn}, pg. 25) 
\begin{equation}
 E(X^r) = \sum^r_{k=0}{r\choose k}\mu^{r-k}\sigma^r E\left(Z^r\right)\notag,
\end{equation}
where $E(Z^r)$ is given by (\ref{eq:mom}).
%Subsequently, the $r$th moment of $X\sim FIN(\mu,\sigma,\alpha,\beta,\kappa)$ is given by (\ref{eq:rawmom}), and the identity $E(X^r)=\sigma^r E(Z^r)$.
%
\subsection{Maximum likelihood estimation}
\label{sec:estim}
The \ac{LL} for a random sample $x_1,x_2,\dots,x_n$ from $X\sim \ac{FIN}(\mu,\sigma,\alpha,\beta,\kappa)$ observations is
\begin{align}\label{eq:LL}
 \text{\ac{LL} }(\mu, \sigma, \alpha, \beta;x_1,x_2,\dots,x_n) & =
 \sum_{i=1}^{n}\text{ln }f(x_i;\mu, \sigma, \alpha, \beta, \kappa),
\end{align}
where $f(\cdot)$ is equivalent to (\ref{eq:lsFIN}). The derivatives of the individual terms in (\ref{eq:LL}) with respect to the \ac{FIN} parameters are given by:
\begin{align*}
\frac{\partial \ac{LL} }{\partial \mu}
&=\left\{\begin{array}{ll}
\frac{\rpsi_2\left(\frac{\alpha}{\beta\kappa},z^{\beta\kappa}\right)\beta\kappa z^{\beta\kappa}}{\sigma z \Gamma\left(\frac{\alpha}{\beta\kappa},z^{\beta\kappa}\right)}
& z \leq 0 \\
-\frac{\rpsi_2\left(\frac{\alpha\kappa}{\beta},\left(\phi^{-\frac{1}{\alpha}} z\right)^{\beta/\kappa}\right)\beta\left(\phi^{-\frac{1}{\alpha}}z\right)^{\beta/\kappa}}{z\sigma\kappa \Gamma\left(\frac{\alpha\kappa}{\beta},\left(\phi^{-\frac{1}{\alpha}} z\right)^{\beta/\kappa}\right)}
& z > 0 \\
\end{array}
\right.
\\
\frac{\partial \ac{LL} }{\partial \sigma}
&=\left\{\begin{array}{ll}
-\frac{\rpsi_2\left(\frac{\alpha}{\beta\kappa},z^{\beta\kappa}\right)\beta\kappa z^{\beta\kappa}}{\sigma z \Gamma\left(\frac{\alpha}{\beta\kappa},z^{\beta\kappa}\right)}-\frac{1}{\sigma}
& z \leq 0 \\
-\frac{\rpsi_2\left(\frac{\alpha\kappa}{\beta},\left(\phi^{-\frac{1}{\alpha}} z\right)^{\beta/\kappa}\right)\beta\left(\phi^{-\frac{1}{\alpha}}z\right)^{\beta/\kappa}}{\sigma\kappa \Gamma\left(\frac{\alpha\kappa}{\beta},\left(\phi^{-\frac{1}{\alpha}} z\right)^{\beta/\kappa}\right)}-\frac{1}{\sigma}
& z > 0 \\
\end{array}
\right.
\end{align*}
\begin{align*}
\frac{\partial \ac{LL} }{\partial \alpha}
&=\left\{\begin{array}{ll}
\frac{\rpsi_1\left(\frac{\alpha}{\beta\kappa},z^{\beta\kappa}\right)}{\beta\kappa\Gamma\left(\frac{\alpha}{\beta\kappa},z^{\beta\kappa}\right)}
    - \frac{\frac{\partial}{\partial\alpha}\left(\delta\right)}{\delta} \\
& z \leq 0 \\
\frac{\frac{\partial}{\partial\alpha}\left(\phi\right)}{\phi}
    +\frac{\rpsi_3\left(\frac{\alpha\kappa}{\beta},\left(\phi^{-\frac{1}{\alpha}}z\right)^{\beta/\kappa}\right)}{\Gamma\left(\frac{\alpha\kappa}{\beta},\left(\phi^{-\frac{1}{\alpha}}z\right)^{\beta/\kappa}\right)}
    - \frac{\frac{\partial}{\partial\alpha}\left(\delta\right)}{\delta}
& z > 0 \\
\end{array}
\right.
\end{align*}
\begin{align*}
\frac{\partial \ac{LL} }{\partial \beta}
&=\left\{\begin{array}{ll}
\frac{\rpsi_3\left(\frac{\alpha}{\beta\kappa},z^{\beta\kappa},\beta\right)}
    {\Gamma\left(\frac{\alpha}{\beta\kappa},z^{\beta\kappa}\right)}
    - \frac{\frac{\partial}{\partial\beta}\left(\delta\right)}{\delta}
& z \leq 0 \\
\frac{\frac{\partial}{\partial\beta}\left(\phi\right)}{\phi}
    + \frac{\rpsi_3\left(\frac{\alpha\kappa}{\beta},\left(\phi^{-\frac{1}{\alpha}}z\right)^{\beta/\kappa},\beta\right)} {\Gamma\left(\frac{\alpha\kappa}{\beta},\left(\phi^{-\frac{1}{\alpha}}z\right)^{\beta/\kappa}\right)}-\frac{\frac{\partial}{\partial\beta}\left(\delta\right)}{\delta}
& z > 0 \\
\end{array}
\right.
\end{align*}
\begin{align*}
\frac{\partial \ac{LL} }{\partial \kappa}
&=\left\{\begin{array}{ll}
-\frac{1}{\kappa} + \frac{\rpsi_3\left(\frac{\alpha}{\beta\kappa},z^{\beta\kappa},\kappa\right)}{\Gamma\left(\frac{\alpha}{\beta\kappa},z^{\beta\kappa}\right)}
    -\frac{\frac{\partial}{\partial\kappa}\left(\delta\right)}{\delta}
& z \leq 0 \\
\frac{\frac{\partial}{\partial\kappa}\left(\phi\right)}{\phi}
    + \frac{1}{\kappa}
    + \frac{\rpsi_3\left(\frac{\alpha\kappa}{\beta},\left(\phi^{-\frac{1}{\alpha}}z\right)^{\beta/\kappa},\kappa\right)}{\Gamma\left(\frac{\alpha\kappa}{\beta},\left(\phi^{-\frac{1}{\alpha}}z\right)^{\beta/\kappa}\right)}
    -\frac{\frac{\partial}{\partial\kappa}\left(\delta\right)}{\delta}
& z > 0 \\
\end{array}
\right.
\end{align*}
where $
 \rpsi_1(u,v)  = \frac{\partial}{\partial u} \Gamma(u,v) = \Gamma \left(u,v\right)\ln{v}+A(u,v),
 \rpsi_2(u,v)   = \frac{\partial}{\partial v} \Gamma(u,v) = -v^{u-1}\text{e}^{-v},
 \rpsi_3(u,v,w) = \rpsi_1(u,v)\frac{\partial}{\partial w} u+ \rpsi_2(u,v)\frac{\partial}{\partial w} v ,
 A(u, v) = G^{3,0}_{2,3} \left(v \left \lvert \begin{gathered} 1, 1 \\ 0, 0, u \end{gathered} \right. \right),
$
$\psi(\cdot)$ is the digamma function, and $G$ is the Meijer's G function; see (\cite{Gr<3dstheyn}, p. 850,902). The remaining values of $
\frac{\partial \phi}{\partial\alpha},
\frac{\partial \phi}{\partial\beta},
\frac{\partial \phi}{\partial\kappa},
\frac{\partial \delta}{\partial\alpha},
\frac{\partial \delta}{\partial\beta},\text{ and}
\frac{\partial \delta}{\partial\kappa}
$ are given by (\ref{eq:llphia}-\ref{eq:lldeltak}) in the Appendix.
\section{Application}
\label{sec:application}
In this section, the \ac{FIN} is applied to publicly available stock returns data. The returns distributions are fitted using \ac{ML} estimation from Section \ref{sec:estim},  and the ghyp package in R \cite{ghypack}, with the simplifying assumption of time-independence of observations. For the validation of marginal distribution fits, we use the in-sample \ac{AIC} \cite{akaike1974new} computed on the data used for estimation and a final verification with out-of-sample \ac{LL}  calculated on the last 20\% of the data time range excluded during estimation. Lower \ac{AIC}  values and higher out-of-sample \ac{LL}  values indicate better fit, respectively. The competitor marginal models are the \ac{HYP}, \ac{GHYP}, \ac{NIG}, and \ac{ST}. These are common models used in portfolio modelling, see \cite{kuchler1999stock} and \cite{Haas2009}. After determining the best-performing marginal distributions using the available performance metrics, the observations are transformed using the fitted \ac{CDF} to be modelled using the copula technique. The copulas used take the following form below:
\[ C_{\Theta }(u_1, u_2,\ldots , u_ m) = \pmb F_\Theta \Bigl (F ^{-1} (u_1), F ^{-1} (u_2),\ldots , F ^{-1} (u_ m)\Bigr ) ,\]
where $\pmb F_\Theta$ is either the multivariate normal or Student’s t distribution function, $\Theta$ the respective correlation and shape parameters, and $F^{-1}$ the quantile function of the chosen marginal distribution.
\subsection{Stock returns data}
\begin{figure}[ht!]
 \centering{
		\includegraphics[width=0.9\linewidth,keepaspectratio]{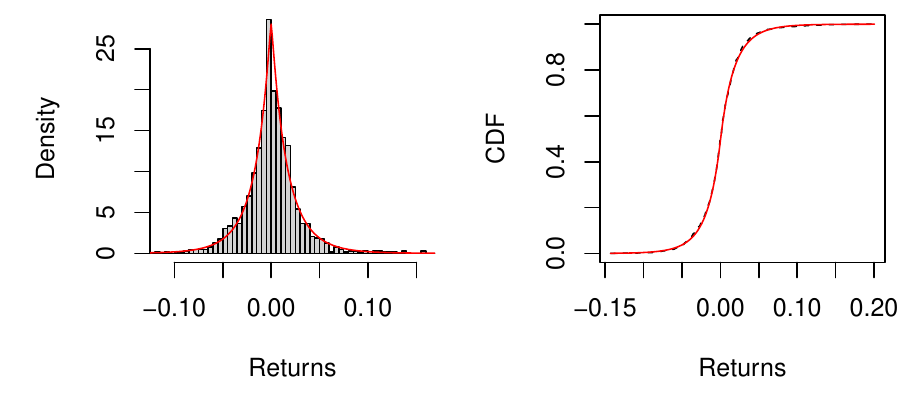}
	}
	\caption{The empirical and fitted \ac{FIN} \ac{PDF} and \ac{CDF} on the \ac{GSIT}  stock returns data.}	\label{fig:gsitfit}
\end{figure}

The data consists of daily log-returns for the shares of five companies listed on the \ac{NASDAQ}, each in different market sectors. The companies are \ac{ACST}, \ac{GSIT}, \ac{LOAN}, \ac{MGI}, and \ac{PSCM}. The period for the data is from 4 January 2016 to 31 December 2020, and it is available at \url{http://finance.yahoo.com}. The summary statistics for the stock returns data are given in Table \ref{tab:5stocksummaries}. It can be seen that the \ac{ACST} and \ac{MGI} data have very large Pearson kurtosis values, making them heavy tailed. There also seems to be skewness present, with \ac{ACST} and \ac{MGI}  again having the largest absolute values. The Spearman rank correlation coefficients of the stock returns data are given in Table \ref{tab:5stockspearman}. Here, it can be seen that \ac{PSCM} is correlated the highest with all the other stocks. An illustration of the shape of these type of returns distributions is given by the histogram and empirical \ac{CDF} in Figure \ref{fig:gsitfit}.
\begin{table}[!htbp] 
\centering 
\begin{tabular}{@{\extracolsep{5pt}} crrrrr} 
\\[-1.8ex]\hline 
\hline \\[-1.8ex]
Stock & n & Avg & Std & Skewness & Kurtosis \\ 
\hline \\[-1.8ex] 
\ac{ACST}  & $1,259$ & $-0.0020$ & $0.081$ & $-1.433$ & $70.575$ \\ 
\ac{GSIT}  & $1,260$ & $0.0010$ & $0.029$ & $0.743$ & $9.117$ \\ 
\ac{LOAN}  & $1,260$ & $0.0004$ & $0.026$ & $-0.325$ & $17.209$ \\ 
\ac{MGI}  & $1,260$ & $-0.0001$ & $0.056$ & $2.787$ & $100.754$ \\ 
\ac{PSCM}  & $1,250$ & $0.0010$ & $0.018$ & $-0.453$ & $8.406$ \\ 
\hline \\[-1.8ex] 
\end{tabular} 
\caption{Summary statistics of stock returns data.} 
\label{tab:5stocksummaries}
\end{table}
\begin{table}[!htbp] \centering 

\begin{tabular}{@{\extracolsep{5pt}} cccccc} 
\\[-1.8ex]\hline 
\hline \\[-1.8ex] 
 & \ac{ACST}  & \ac{GSIT}  & \ac{LOAN}  & \ac{MGI}  & \ac{PSCM}  \\ 
\hline \\[-1.8ex] 
\ac{ACST}  & $1$ & $0.048$ & $0.004$ & $0.097$ & $0.152$ \\ 
\ac{GSIT}  & $0.048$ & $1$ & $0.080$ & $0.164$ & $0.322$ \\ 
\ac{LOAN}  & $0.004$ & $0.080$ & $1$ & $0.056$ & $0.167$ \\ 
\ac{MGI}  & $0.097$ & $0.164$ & $0.056$ & $1$ & $0.272$ \\ 
\ac{PSCM}  & $0.152$ & $0.322$ & $0.167$ & $0.272$ & $1$ \\ 
\hline \\[-1.8ex] 
\end{tabular} 
\caption{Spearman rank correlation coefficients for stock returns data.} 
\label{tab:5stockspearman} 
\end{table}
\subsection{Modelling}
In Table \ref{tab:5stockinsample}, the in-sample \ac{AIC}  statistics are given and shows that the fitted \ac{FIN} has the lowest value for all stocks. In Table \ref{tab:5stockoutofsample}, the out-of-sample \ac{LL}  is given and shows that the fitted \ac{FIN} has the highest value for all stocks. It can therefore be concluded that the \ac{FIN} is the overall best-fitting model for the marginal distributions. In Table \ref{tab:5stockcopfit}, the in-sample \ac{AIC}  and out-of-sample \ac{LL}  is given for the fitted copulas. Since the student t copula has the lowest \ac{AIC}  and highest \ac{LL} , it can therefore be concluded that the t copula fits the dependence structure best. The estimated correlation parameters is given in Table \ref{5stockcorrest} and the fitted degrees of freedom for the student t copula is $16.481$. 
\begin{table}[!htbp] \centering 
%\fontsize{8}{0}
\begin{tabular}{cccccc}
\\[-1.8ex]\hline 
\hline \\[-1.8ex] 
Stock &  \ac{GHYP}  &  \ac{HYP}  &  \ac{NIG}  & \ac{ST} & \ac{FIN} \\ 
\hline \\[-1.8ex] 
\ac{ACST}  & $-3,402.034$ & $-3,279.898$ & $-3,392.053$ & $-3,403.758$ & $-3,729.600$ \\ 
\ac{GSIT}  & $-4,887.852$ & $-4,889.277$ & $-4,888.362$ & $-4,882.220$ & $-4,889.489$ \\ 
\ac{LOAN}  & $-5,368.226$ & $-5,370.473$ & $-5,368.627$ & $-5,369.186$ & $-5,510.395$ \\ 
\ac{MGI}  & $-4,116.699$ & $-4,024.419$ & $-4,118.700$ & $-4,116.190$ & $-4,142.729$ \\ 
\ac{PSCM}  & $-5,857.888$ & $-5,857.855$ & $-5,854.014$ & $-5,849.009$ & $-5,863.092$ \\ 
\hline \\[-1.8ex] 
\end{tabular} 
\caption{Fitted distributions and their in-sample \ac{AIC}  statistics for the stock returns data.} 
\label{tab:5stockinsample} 
\end{table} 
\begin{table}[!htbp] 
\centering
\begin{tabular}{ccclll} 
\\[-1.8ex]\hline 
\hline \\[-1.8ex] 
Stock &  \ac{GHYP}  &  \ac{HYP}  &  \ac{NIG}  & \ac{ST} & \ac{FIN} \\ 
\hline \\[-1.8ex]
\ac{ACST}  & 261.267 & 245.05 & 264.875 & 260.811 & 379.193 \\ 
\ac{GSIT}  & 387.432 & 383.528 & 387.64 & 385.538 & 387.98 \\ 
\ac{LOAN}  & 447.483 & 414.289 & 440.541 & 450.07 & 458.919 \\ 
\ac{MGI}  & 288.557 & 291.974 & 288.657 & 283.968 & 291.994 \\ 
\ac{PSCM}  & 440.12 & 449.19 & 446.261 & 449.142 & 453.623 \\ 
\hline \\[-1.8ex] 
\end{tabular} 
\caption{Out-sample \ac{LL}  values for stock returns data} 
\label{tab:5stockoutofsample} 
\end{table} 
\begin{table}[!htbp] 
\centering 
\begin{tabular}{@{\extracolsep{5pt}} ccc} 
\\[-1.8ex]\hline 
\hline \\[-1.8ex] 
Copula & \ac{AIC}  & \ac{LL}  \\ 
\hline \\[-1.8ex] 
Normal & $-185.853$ & $60.646$ \\ 
Student-t & $-200.995$ & $93.742$ \\ 
\hline \\[-1.8ex] 
\end{tabular}
\caption{In-sample \ac{AIC}  and out-of-sample \ac{LL}  is for the fitted copula functions.}
\label{tab:5stockcopfit} 
\end{table}
\begin{table}[!htbp] \centering 
\begin{tabular}{@{\extracolsep{5pt}} cccccc} 
\\[-1.8ex]\hline 
\hline \\[-1.8ex] 
 & \ac{ACST}  & \ac{GSIT}  & \ac{LOAN}  & \ac{MGI}  & \ac{PSCM}  \\ 
\hline \\[-1.8ex] 
\ac{ACST}  & $1$ & $0.064$ & $-0.013$ & $0.091$ & $0.184$ \\ 
\ac{GSIT}  & $0.064$ & $1$ & $0.039$ & $0.219$ & $0.369$ \\ 
\ac{LOAN}  & $-0.013$ & $0.039$ & $1$ & $0.021$ & $0.096$ \\ 
\ac{MGI}  & $0.091$ & $0.219$ & $0.021$ & $1$ & $0.321$ \\ 
\ac{PSCM}  & $0.184$ & $0.369$ & $0.096$ & $0.321$ & $1$ \\ 
\hline \\[-1.8ex] 
\end{tabular} 
\caption{Fitted correlation estimates for stock returns data.}
\label{5stockcorrest} 
\end{table}
\subsection{Application conclusion}
 Based on the modelling results presented in Tables \ref{tab:5stockinsample}, \ref{tab:5stockoutofsample}, and \ref{tab:5stockcopfit} it can be concluded that the newly proposed \ac{FIN} model is the best-fitting model for the marginal distributions, while the student t copula fits the dependence structure best, with a fitted degrees of freedom of 16.481. This analysis reveals that the \ac{FIN} and t copula models are well suited to the given data set, providing an accurate representation of the underlying marginal and dependence structures.
\section{Concluding remarks}
This chapter focuses on the analysis of characteristics such as body shape, skewness, and kurtosis that can be observed in different data sets. In particular, we explore the generalisations of the \ac{EP} distribution for this purpose. Through an examination of existing skew \ac{EP} distributions and skewing mechanisms, we identify the advantages and drawbacks of previous literature in this regard. In order to improve on this, we introduce the derivative kernel skewing paradigm to derive the \ac{FIN} distribution. The \ac{FIN} distribution has several desirable properties, including interpretable parameters, finite moments, tractable equations, a ``joint" free mode, a mode equal to the location parameter, and kurtosis skewing similar to the Jones skew t and Azzalini skew normal. We also provide statistical properties such as the \ac{PDF}, \ac{CDF}, moments, and \ac{ML} equations. To demonstrate the modelling performance of the \ac{FIN} distribution, we apply it to stock returns data from five companies listed on the \ac{NASDAQ}. Our results show that the \ac{FIN} distribution achieves the best-fitting marginal distributions according to in-sample and out-of-sample metrics, while the dependence structure is modelled best with the t copula according to the sample metrics. In conclusion, the \ac{FIN} distribution is a flexible model without sacrificing interpretability.
\section*{Acknowledgements}
We would like to express our sincere gratitude to the two anonymous reviewers for their invaluable feedback and comments that significantly improved the presentation of the work. This work is based on the research supported in part by the National Research Foundation of South Africa (Ref.: SRUG2204203965; RA171022270376, UID:119109; RA211204653274, Grant No. 151035), as well as the Centre of Excellence in Mathematical and Statistical Sciences at the University of the Witwatersrand. The work of the author Muhammad Arashi is supported by the Iran National Science Foundation (INSF) Grant No. 4015320. Opinions expressed and conclusions arrived at are those of the authors and are not necessarily to be attributed to the funders of this work.
\section{Appendix}
\addcontentsline{toc}{section}{Appendix}
\subsection*{Lemmas}
\label{asec:A}
\subsubsection*{Lemma 1}
 Let $b,c>0$. Then the following limit holds true
\begin{equation}
  \label{eq:lem1}
  \lim\limits_{x\to \infty} x^{a}\Gamma\left( b,x^c\right) =0, \forall a\in\mathbb{R}.
 \end{equation}
\subsubsection*{Proof}
 If $a\le0$ both factors on the left-hand side of (\ref{eq:lem1}) tend to zero as $x$ tends to infinity.
 If $a>0$, by L'Hospital rule
\begin{equation}
  \lim\limits_{x\to \infty} x^a\Gamma\left( b,x^c \right)
  =
  \lim\limits_{x\to \infty}
  \frac{x^{a+b+c-1}}{\text{e}^{x^c}}
  \frac{c}{a}
  = 0\notag.
 \end{equation}
\subsubsection*{Lemma 2}
Let $a>-1$, and $b,c>0$, then the following integral identity holds true
\begin{equation}
  \label{eq:lem2}
  \int_x^\infty t^a\Gamma\left( b,t^c\right)dt
  = \frac
  {\Gamma\left(\frac{a+bc+1}{c},x^c \right)
  -x^{a+1}\Gamma\left(b,x^c \right)}
  {a+1}.
 \end{equation}
 \subsubsection*{Proof}
Let $y=t^c$, that implies $t=y^{\frac{1}{c}}$. Integrating by parts, where
 $v'(y)=\frac{a+1}{c}y^{\frac{a+1}{c}-1}$ and $u(y)=\Gamma\left(b,y \right)$. The latter implies that $v(y)=y^{\frac{a+1}{c}}$ and $u'(y)=-y^{b-1}\text{e}^{-y}$. The integral is evaluated as
\begin{gather}
  \int^{\infty}_{x}t^r\Gamma\left(b,t^c \right)dt
  =
  (a+1)^{-1}
  \left. y^{\frac{a+1}{c}}\Gamma\left(b,y \right)\right|^\infty_{x^c}
  -(a+1)^{-1}
  \int^{\infty}_{x^c}y^{\frac{a+1}{c}}
  \left( -y^{b-1}\text{e}^{-y}\right) dy.\notag
 \end{gather}
  Noting (\ref{eq:lem1}) that $\lim\limits_{x\to \infty} y^{\frac{a+1}{c}}\Gamma\left(b,y \right)=0$
 the result follows.
\subsubsection*{Lemma 3}
Let $a>-1$, and $b,c>0$. Then the following integral identity holds true
\begin{equation}
  \label{eq:lem3}
  \int_0^\infty x^a\Gamma\left(b,x^c\right)dx
  = \frac
  {\Gamma\left(\frac{a+bc+1}{c}\right)}
  {a+1}.
 \end{equation}
 \subsubsection*{Proof}
The result follows by evaluating the limit, $\lim_{x \to 0^+}$, over the integral in (\ref{eq:lem2}).
\subsubsection*{Lemma 4}
Let $r>-1$ and $k(\cdot)$ be the kernel of the {FIN} distribution (\ref{eq:FINkern}). Then the following integral identity holds true
\begin{equation}
    \int^{\infty}_{-\infty}{z^r k(z) dz}=
    (-1)^r\frac{ \kappa^{-1}}{r+1} \Gamma \left( \frac{r + \alpha + 1}{\beta \kappa} \right) 
    + \phi^{\frac{r + \alpha + 1}{\alpha}}\frac{\kappa}{r+1} 
    \Gamma \left( \frac{r + \alpha + 1}{\beta}\kappa \right).\label{eq:lem4}\notag
\end{equation}
\subsubsection*{Proof}
Substituting (\ref{eq:FINkern}) and defining the transformations $s = -z$, and $t = \phi^{-\frac{1}{\alpha}} z$. The integral is evaluated as 
\begin{equation}
    \int^{\infty}_{-\infty}{z^r k(z) dz}=\kappa^{-1} (-1)^{r} \int^{\infty}_{0}{s^r \Gamma \left( \frac{\alpha}{\beta \kappa}, s^{\beta \kappa} \right) ds} + \phi^{\frac{\alpha + r + 1}{\alpha}} \kappa \int^{\infty}_{0}{t^r \Gamma \left( \frac{\alpha \kappa}{\beta}, t^{\beta / \kappa} \right) dt}.\label{eq:lem4first}\notag
\end{equation}
 Noting from (\ref{eq:lem3}) that $\int_0^\infty x^a\Gamma\left(b,x^c\right)dx
  = \frac
  {\Gamma\left(\frac{a+bc+1}{c}\right)}
  {a+1}$ the result follows.
\subsection{Maximum likelihood estimation}
The remaining descriptive values for maximum likelihood partial derivatives in Section \ref{sec:estim} is given by:
\begin{align}
\frac{\partial}{\partial \alpha}\phi &=
    \frac{\Gamma \left(\frac{\alpha}{\beta \kappa}\right)\psi\left(\frac{\alpha}{\beta \kappa}\right)\left( \frac{1}{\beta \kappa}\right) -\psi\left(\frac{\alpha \kappa}{\beta }\right)\left( \frac{\kappa}{\beta} \right)\Gamma \left(\frac{\alpha}{\beta \kappa}\right)}{\kappa^{2}\Gamma \left(\frac{\alpha \kappa}{\beta}\right)}\label{eq:llphia}
\\
\frac{\partial}{\partial \beta}\phi 
    & = \frac{\Gamma \left (\frac{\alpha}{\beta \kappa}\right )\psi\left(\frac{\alpha}{\beta \kappa}\right)\left( -\frac{\alpha}{\beta^{2} \kappa}\right) -\psi\left(\frac{\alpha \kappa}{\beta }\right)\left( -\frac{\alpha\kappa}{\beta^{2} }\right)\Gamma \left (\frac{\alpha}{\beta \kappa}\right )}{\kappa^{2}\Gamma \left (\frac{\alpha \kappa}{\beta}\right )}\label{eq:llphib}
\\
\frac{\partial}{\partial \kappa}\phi & = \frac{\Gamma \left (\frac{\alpha}{\beta \kappa}\right )\left (\alpha\psi\left(\frac{\alpha}{\beta \kappa}\right)+ \kappa \left(\alpha \kappa \psi\left(\frac{\alpha \kappa}{\beta }\right) +2\beta \right )\right)}{\beta\kappa^{4}\Gamma \left (\frac{\alpha \kappa}{\beta}\right )^{2}}\label{eq:llphik}
\end{align}
\begin{align}
\frac{\partial}{\partial \alpha} \delta 
    & = \frac{\Gamma \left( \frac{\alpha +1}{\beta \kappa} \right)\psi \left( \frac{\alpha +1}{\beta \kappa} \right)}{\beta\kappa^{2}}+\frac{\phi^{\frac{1}{\alpha}}\left( \alpha \left(\alpha+1 \right) \frac{\partial}{\partial \alpha}\left(\phi\right) -\phi \ln{\phi}\right)}{\alpha^{2}}\kappa\Gamma \left( \frac{\alpha+1}{\beta}\kappa\right)\nonumber\\&\quad 
    + \frac{\phi^{\frac{\alpha+1}{\alpha}}\kappa^{2} \Gamma \left( \frac{\alpha+1}{\beta}\kappa\right)\psi \left( \frac{\alpha+1}{\beta}\kappa\right)}{\beta}\label{eq:lldeltaa}
\\
\frac{\partial}{\partial\beta}\delta 
    & = - \frac{
    \left(\alpha+1\right) \Gamma\left(\frac{\alpha+1}{\beta\kappa}\right) \psi\left(\frac{\alpha +1}{\beta\kappa}\right)
    }
    { \beta^{2}\kappa^{2} }
    - \frac{
    \left(\alpha+1\right) \phi^{\frac{\alpha+1}{\alpha}} \kappa^{2} \Gamma\left(\frac{\alpha+1}{\beta}\kappa\right) \psi\left(\frac{\alpha+1}{\beta}\kappa\right) }
    { \beta^{2} } \nonumber \\
    & \quad
    + \frac{\left(\alpha+1\right)\phi^{\frac{1}{\alpha}}\left(\phi\right)}{\alpha}\kappa\Gamma\left(\frac{\alpha+1}{\beta}\kappa\right)\label{eq:lldeltab}
\\
\frac{\partial}{\partial \kappa} \delta 
    & = -\frac{\Gamma \left ( \frac{\alpha+1 }{\beta\kappa}\right )}{\kappa^{2}}
    -\frac{\left(\alpha +1 \right )\Gamma \left ( \frac{\alpha +1}{\beta \kappa} \right )\psi \left ( \frac{\alpha +1}{\beta \kappa} \right )}{\beta\kappa^{3}}
    +\frac{\left(\alpha+1\right) \phi^{\left(\frac{\alpha+1}{\alpha}\right)} \kappa \Gamma\left(\frac{\alpha+1}{\beta}\kappa\right) \psi \left(\frac{\alpha+1}{\beta}\kappa\right)}
    {\beta} \nonumber\\
    & \quad
    + \frac{\left(\alpha+1\right)\phi^{\left(\frac{1}{\alpha}\right)} \left(\phi\right) } {\alpha}\kappa\Gamma\left(\frac{\alpha+1}{\beta}\kappa\right)
    +\phi^{\frac{\alpha+1}{\alpha}}\Gamma\left(\frac{\alpha+1}{\beta}\kappa\right)\label{eq:lldeltak}
\end{align}

\bibliography{btnormalwork, Finref, Skewref, GG}
\bibliographystyle{abbrv}
\end{document}